# Casimir effect for quantum graphs

D.U. Matrasulov, J.R.Yusupov, P.K.Khabibullaev, A.A.Saidov
Heat Physics Department of the Uzbek Academy of Sciences,
28 Katartal St., 700135 Tashkent, Uzbekistan

Abstract

The Casimir effects for one-dimensional fractal networks, so-called quantum graphs is studied. Based on the Green function approach for quantum graphs zero-point energy for some simplest topologies is written explicitly.

The Casimir effect is a manifestation of the zero-point oscillations of the vacuum of a quantum (electromagnetic, scalar, etc.) field [1-6]. In the simplest case it is the interaction of two neutral, parallel perfectly conducting plates. This zero-point energy can be calculated via solving quantum field theoretical equations with given boundary conditions. Despite the fact that almost sixty years have passed since from H.B.G. Casimir published his famous paper, Casimir effect is still remaining as a hot topic in contemporary physics. Especially, Casimir effect has gained new development due to recent progress made in nanoscale physics, such as nanomechanics and mesoscopic physics.

In this paper, we treat Casimir effect on quantum graphs. Namely, using the Green function approach for quantum graphs we explore possibility of computations of zero-point energy in such systems. Quantum graphs are an important class of models, which have proven to be valuable for understanding quantum phenomena in mesoscopic and disordered systems as well as problems in quantum information and molecular electronics [7-9]. These are one-dimensional networks, which were introduced earlier in quantum chemistry to model free electron motion in organic molecules. During the last decade they found application as models in solid state physics, acoustics, mesoscopic physics and quantum information theory as well as for organic molecules being the key building blocks for a possible new generation of electronic devices on molecular scales. Quantum graphs have been shown to serve as accurate models for the study of quantum transport and spectral statistics [6,9] in nanosized systems. Recently quantum graphs are studied experimentally by using microwave networks consisting of coaxial waveguides [10]. Zero point energy in quantum graphs may play important role in various systems (e.g., polymers, molecular networks, microwave networks and other supramolecular structures), whose dynamics can be modelled by quantum graphs, as well as in nanomechanics.

We give first brief definition of quantum graph.

Graph consists of $V$ vertices connected by $B$ bonds. The valency of a vertex is the number of bonds meeting at that vertex. When the vertices $i$ and $j$ are connected, we denote the connecting bond by $b = (i,j)$.

Particle dynamics in a quantum graph is described by the following one-dimensional Schrödinger [6] (in units $\hbar = 2m = 1$):

$$\left(-i\frac{d}{dx} - A_b\right)^2 \Psi_b(x) = k^2 \Psi_b(x), \quad b = (i,j), \tag{1}$$

where on each bond $b$, the component $\Psi_b$ of the total wavefunction $\Psi$ is a solution of the eq.(1). Here a "magnetic vector potential" $A_b$ is introduced to break time-reversal symmetry. The wavefunction, $\Psi_b$ satisfies boundary conditions at the vertices, which ensure continuity and curent conservation [6]. The boundary conditions are given as follows [6]: for every $i = 1, ..., V$

$$\begin{cases} \bullet \text{ Continuity,} \\ \Psi_{i,j}(x)\big|_{x=0} = \varphi_i, \quad \Psi_{i,j}(x)\big|_{x=L_{i,j}} = \varphi_j, \quad \text{for all } i < j \text{ and } C_{i,j} \neq 0 \\ \bullet \text{ Current conservation,} \\ \sum_{j<i} C_{i,j}\left(iA_{j,i} - \frac{d}{dx}\right)\Psi_{j,i}(x)\bigg|_{x=L_{i,j}} + \sum_{j>i} C_{i,j}\left(-iA_{i,j} + \frac{d}{dx}\right)\Psi_{i,j}(x)\bigg|_{x=0} = \lambda_i \varphi_i \end{cases} \tag{2}$$

The parameters $\lambda_i$ are free parameters which determine the type of boundary conditions. The special case of zero $\lambda_i$'s corresponds to Neumann boundary conditions. Dirichlet boundary conditions correspond to the case when all the $\lambda_i = \infty$.

The eigenfunctions of the graph are completely determined by the set of functions $\{\varphi_i\}_{i=1}^V$, values of the wavefunction at the vertices. At any bond $b = (i,j)$ the component $\Psi_b$ can be written in terms of its values on the vertices $i$ and $j$ as

$$\Psi_{i,j} = \frac{e^{iA_{i,j}x}}{\sin kL_{i,j}}(\varphi_i \sin[k(L_{i,j} - x)]) + \varphi_j e^{iA_{i,j}L_{i,j}} \sin kx)C_{i,j}, \quad i < j. \tag{3}$$

In the case of Dirichlet boundary conditions the eigenfunction has simple structure:

$$\Psi_b = \frac{e^{iA_b x}}{\sqrt{L_b}} \sin\left(\frac{n_b \pi x}{L_b}\right)$$

and the eigenvalues are given by

$$k_n^{(b)} = \frac{n_b \pi}{L_b} \quad \text{for all} \quad n_b > 0.$$

Several approaches are used to treat of such properties of quantum graphs as spectral statistics, transport and other properties. In particular, periodic orbit theory is used to describe spectral statistics in semiclassical approach [6]. Scattering approach allows to treat particle transport in quantum graphs [6]. Recently the Green function approach is developed for exploring general quantum graphs [8].

Green function of a quantum graph, is defined by the equations

$$\frac{d^2}{dx^2}G_1(x,x') + k^2 G_1(x,x') = \frac{2m}{\hbar^2}\delta(x-x'), \quad 0 < x, x' < l_1$$

$$\frac{d^2}{dx^2}G_{\hat{1}}(x,x') + k^2 G_{\hat{1}}(x,x') = \frac{2m}{\hbar^2}\delta(x - l_1 + x'), \quad 0 < x, x' < l_{\hat{1}} = l_1$$

$$\frac{d^2}{dx^2}G_b(x,x') + k^2 G_b(x,x') = 0, \quad \forall b \neq \{1,\hat{1}\}$$

Note that $x'$ is fixed to belong to the bond 1. The Green function satisfies the same boundary condition as the wave function. Recently the explicit form of the Green function for some graphs was derived and a prescription for finding the exact Green function for general (open or closed) graphs was developed [8].

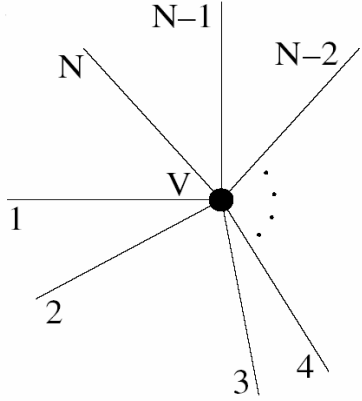
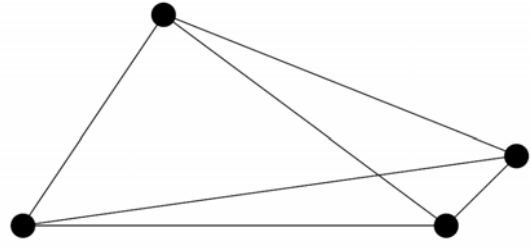

Figure 1.  Figure 2.

For the graph which is given in Fig 1. the exact Green function is given by

$$G_{nl} = \frac{1}{2ik}\left\{\delta_{nl}\exp[ik|x_f - x_i|] + S_{nl}(k)\exp[ik(x_f + x_i)]\right\}$$

where $S_{nl}(k)$ is the scattering matrix defined from the equation

$$\psi_n(x;k) = \delta_{nj}\exp[-ikx] + S_{nj}(k)\exp[ikx].$$

This scattering matrix satisfies the following conditions

$$SS^+ = S^+ S = 1, \quad S^+(k) = S(-k),$$

and can be related to the transmission and reflection coefficients, $T_{n,j}$ and $R_n$ as follows:

$$S_{nn} = R_n \quad S_{nj} = T_{nj}$$

For the graph in Fig. 2 which has the form of tetrahedron, the Green function can be written as

$$G = \frac{1}{2ikg}\{(1 - \mathfrak{I}_1 \exp[ik\ell])\exp[ik(x_f - x_i)] + \mathfrak{R}_1 \exp[ik(x_f + x_i)] + \qquad (4)$$
$$+ (\mathfrak{I}_2 + (\mathfrak{R}_1\mathfrak{R}_2 - \mathfrak{I}_1\mathfrak{I}_2)\exp[ik\ell])\exp[ik(l - x_f + x_i)] +$$
$$+ \mathfrak{R}_2 \exp[ik(2l - x_f - x_i)]\}.$$

$R = (\gamma - (N-2)ik)/(Nik - \gamma)$  $(T = 2ik/(Nik - \gamma))$, where $N = 3$

$\mathfrak{I}_1 = \mathfrak{I}_2 = \mathfrak{I}$  $\mathfrak{R}_1 = \mathfrak{R}_2 = \mathfrak{R}$

$\mathfrak{I} = T^2 \exp[2ik\ell]\{(R+T)(1 - R \cdot \exp[ik\ell]) + 2T^2 \exp[ik\ell]\}/f$, $\mathfrak{R} = -\{R - R^2 \exp[ik\ell] +$
$(T^3 - 2R^2T - R^3)\exp[2ik\ell] + (R^4 + 2R^3T - 2R^2T^2 - 3RT^3 + 2T^4)\exp[3ik\ell]\}/f$
$f = 2ik[1 - R \cdot \exp[ik\ell] - (R+T)^2 \exp[2ik\ell] - (2T^3 + RT^2 - 2R^2T - R^3)\exp[3ik\ell]]$.

In the Green function based approach the Casimir energy is calculated as [2,8]

$$E_C = \lim_{\tau \to 0} i\frac{\partial^2}{\partial \tau^2} \int_{-\infty}^{\infty} dk \int_0^L \Gamma(\mathbf{x}, \mathbf{x}, \tau) dx \qquad (5)$$

where $\Gamma(\vec{x}, \vec{x}, \tau)$ is the inhomogeneous part of the Green function which is given by

$$G(\vec{x}, \vec{x}', x - x_0) = G_0(x - x') + \Gamma(\vec{x}, \vec{x}, x - x_0)$$

The function $\Gamma(\vec{x}, \vec{x}, \tau)$ is defined by given boundary conditions, geometry or topology of the given configuration. The parameter $L$ defines the size characteristics of the given geometry (in the case of parallel plates $L$ is the distance between the plates). In the case of quantum graph $L$ is the length of a bond. Inserting the Green function in eq. (4) into the eq. (5) and calculating the obtained integral numerically we get the Casimir energy for the graph. The integral over $x$ can be taken easily. In particular, for the graph given by Fig. 2 we have

$$E = \lim_{\tau \to 0} \frac{\partial^2}{\partial \tau^2} \int_{-\infty}^{\infty} -\frac{[(1 + (\mathfrak{R}^2 - \mathfrak{I}^2)e^{2ikl})ikl + (e^{2ikl} - 1)\mathfrak{R}]}{2k^2 g} e^{ik\tau} dk \qquad (6)$$

The dependence of the Casimir energy of the lentgh of the bond (here for the simplicity we assume that all the bond have the same length, which is equal to $L$) can be obtained by integrating numerically the eq.(6).

Thus, we have developed a prescription that allows to calculate zero-point energy for quantum graphs. It should be noted that within this prescription we cannot speak about Casimir force since the graphs considered within our approach are one-dimensional networks.

This work is supported in part by the grants of Foundation of Support of Fundamental Research of the UzAS (Ref.Nr.70-06) and by the grant of the CCSTD (F-2-1-1).